# Molecular dynamics of arbitrarily shaped granular particles

Thorsten Pöschel[*]

*Arbeitsgruppe Nichtlineare Dynamik, Universität Potsdam, Am Neuen*

*Palais, D–14415 Potsdam, Germany.*

Volkhard Buchholtz[†]

*Humboldt–Universität zu Berlin, Institut für Physik, Unter den Linden 6,*

*D–10099 Berlin, Germany*

(August 1, 1995)

## Abstract

We propose a new model for the description of complex granular particles and their interaction in molecular dynamics simulations of granular material in two dimensions. The grains are composed of triangles which are connected by deformable beams. Particles are allowed to be convex or concave. We present first results of simulations using this particle model.

## I. INTRODUCTION

Molecular dynamics has been proven to be a well suited method to investigate the dynamic and static behavior of granular matter. The concept of molecular dynamics was initially used to simulate the dynamics of atoms and molecules, i.e. of particles without inner degrees of freedom. Alder and Wainwright who belong to the inventors of this method investigated already 1957 numerically the low density approximation of the entropy $S = -N\,k_B\,\langle \ln\,f(p,t)\rangle$ ($f$ is the one–particle probability density in momentum space) in a hard sphere system consisting of $N = 100$ particles [1]. There are applications of molecular dynamics in many fields of interest and molecular dynamics became one of the standard methods in computational physics. Recent molecular dynamics simulations are of very high algorithmic complexity, and systems consisting of up to $10^9$ particles (with simple interaction forces) have been simulated (see e.g. [2–4]). A very interesting overview on the historical development is given in the first part of Hoover's book [5].

---

[*]Email: thorsten@hlrsun.hlrz.kfa–juelich.de

[†]Email: volkhard@itp02.physik.hu–berlin.de

In molecular dynamics of granular particles, sometimes called "granular dynamics", each of the "molecules" is a macroscopic body typically with a diameter of the order $D \approx 0.1\ldots 1\ mm$ which has its own thermodynamic properties, i.e. it has internal degrees of freedom and hence it can dissipate mechanical energy when it is subjected to outer forces. In some special cases as in simulations of planetary rings [6,7] the diameter can also be of the order $D \approx 1\ mm \ldots 10\ m$. In the most simple case one can express the inelastic nature of the collisions of the grains by introducing restitution coefficients which describe the relative velocities in normal and tangential direction $\vec{v}_{ij}^{N}$ and $\vec{v}_{ij}^{T}$ of two granular particles $i$ and $j$ after a collision as functions of the relative velocities before the collision $\vec{V}_{ij}^{N}$ and $\vec{V}_{ij}^{T}$

$$\vec{v}_{ij}^{N} = -\epsilon_N \cdot \vec{V}_{ij}^{N} \quad (0 \leq \epsilon_N \leq 1) \tag{1a}$$

$$\vec{v}_{ij}^{T} = -\epsilon_T \cdot \vec{V}_{ij}^{T} \quad (-1 \leq \epsilon_T \leq 1)\ . \tag{1b}$$

Here the relative velocity is given by

$$\vec{v}_{ij} = \dot{\vec{r}}_i - \dot{\vec{r}}_j \tag{2}$$

or, including the particle rotation with angular velocities $\vec{\omega}_i$ and $\vec{\omega}_j$ , by

$$\vec{v}_{ij} = \left(\dot{\vec{r}}_i + \omega_i \times R_i\left(-\vec{n}\right)\right) - \left(\dot{\vec{r}}_j + \omega_j \times R_j \vec{n}\right) \tag{3}$$

with the unity vectors

$$\vec{n} = \frac{\vec{r}_i - \vec{r}_j}{|\vec{r}_i - \vec{r}_j|} \tag{4a}$$

$$\vec{t} = \begin{pmatrix} 0 & -1 \\ 1 & 0 \end{pmatrix} \cdot \frac{\vec{r}_i - \vec{r}_j}{|\vec{r}_i - \vec{r}_j|} \ . \tag{4b}$$

Hence the relative velocities at the point of contact are

$$\vec{v}_{ij}^N = \vec{n}\left(\vec{v}_{ij} \cdot \vec{n}\right) \tag{5a}$$

$$\vec{v}_{ij}^T = \vec{t}\left(\vec{v}_{ij} \cdot \vec{t}\right) \ . \tag{5b}$$

In general the normal and tangential restitution coefficients $\epsilon_N$ and $\epsilon_T$ are themselves functions of the relative impact velocities $\epsilon_N = \epsilon_N\left(V_{ij}^N, V_{ij}^T\right)$ and $\epsilon_T = \epsilon_T\left(V_{ij}^N, V_{ij}^T\right)$. For the case that the particles are rough homogeneous spheres the coefficients $\epsilon_N$ and $\epsilon_T$ can be calculated analytically [8] solving the viscoelastic equations of the particles. Using this model one neglects the complicated interaction between particles during the collision. The approximation is justified if collisions are relatively rare events, i.e. if the mean free path between collisions is long and one can restrict the calculation to two–particle–interactions only. If a system is in this regime one calls it a "granular gas". Event driven algorithms have been applied in many papers, recently e.g. in [9,10] for an investigation of inelastic clustering of granular particles (using constant restitution coefficients) and by Luding et al. to simulate vertically shaken material [11,12]. For sophisticated implementations of such algorithms see [13,14].

In many cases of interest, however, the particles collide very often or they touch each other permanently, i.e. the system reveals static properties. Examples for such systems are flows through pipes and hoppers, ball mills, shaken materials, sand heaps and many others. In the model proposed by Cundall and Strack [15] and Haff and Werner [16] the particles feel restoring forces in normal and tangential direction, and they lose kinetic energy during the collisions. If the particles $i$ and $j$ at the positions $\vec{r}_i$ and $\vec{r}_j$ with radiuses $R_i$ and $R_j$ touch each other they feel the force

$$\vec{F}_{ij} = F_{ij}^N \vec{n} + F_{ij}^T \vec{t} \tag{6a}$$

with

$$F_{ij}^N = Y \cdot (R_i + R_j - |\vec{r}_i - \vec{r}_j|)^{\frac{3}{2}} - m_{ij}^{eff} \cdot \gamma_N \cdot (\dot{\vec{r}}_i - \dot{\vec{r}}_j) \cdot \vec{n} \qquad (6b)$$

$$F_{ij}^T = \text{sign}(v_{ij}^{rel}) \cdot \min\left(m_{ij}^{eff}\gamma_T \left|v_{ij}^{rel}\right|, \mu \left|F_{ij}^N\right|\right) \qquad (6c)$$

$$v_{ij}^{rel} = (\dot{\vec{r}}_i - \dot{\vec{r}}_j) \cdot \vec{t} + R_i \cdot \Omega_i + R_j \cdot \Omega_j \qquad (6d)$$

$$m_{ij}^{eff} = \frac{m_i \cdot m_j}{m_i + m_j} \ . \qquad (6e)$$

$Y$ is the Young modulus, $\gamma_N$ and $\gamma_T$ are the damping coefficients in normal and tangential direction and $\mu$ is the Coulomb friction coefficient. Eq. (6d) describes the relative velocity of the surfaces of the particles at the point of contact and eq. (6e) gives the effective mass. Eq. (6c) takes the Coulomb friction law into account, saying that two particles slide on top of each other if the shear force overcomes $\mu$ times the normal force. Eq. (6b) comes from the Hertz law [17] for the force between two rigid spheres which are in contact with each other.

This model has been successfully applied to simulate the behavior of dry granular material in many works describing various physical phenomena. It developed to the standard method for calculations of granular dynamics. Some of these problems are size segregation and convection in vibrated material in two and three dimensions (e.g. [18–21]), the flow in hoppers (e.g. [22–24]) and pipes (e.g. [25,26]), the flow in rotating cylinders (e.g. [27–29]), the motion of granular material on an inclined surface [30,31], sound propagation in granular material [32], the onset of turbulence [33,34] and many others. Many experimental results of many authors could be reproduced numerically using this type of molecular dynamics simulations. There have been developed very efficient algorithms for the case of short range interaction with large force gradients as it is typical for granular materials, e.g. [35–37]. The model by Cundall and Strack yields reliable results in more dynamic systems, i.e. when the static friction of the particles does not play a major role. When the static properties of the material govern the behavior, however, the model might fail. In [38] is shown that one cannot build up a stable sand heap with a finite inclination of such particles but the heap dissolves under its own mass and the angle becomes smaller with rising number of particles. Lee [39] simulated static friction effects using spheres which are connected by springs. When two particles touch each other for a certain time a spring "grows" between these particles and keeps the particles together when forces are applied which would separate them. When the separating forces become too large the springs break and the particles move freely. Using this model Lee reproduced the finite angle of repose in a sand heap, however, the growing and breaking springs have no direct equivalent in nature, and hence the model seems to be quite artificial.

Another simulation method which corresponds to the previous one for the case of very large damping has been introduced by Visscher and Bolsterli [40]. In each iteration step all of the grains are moved according to the motion of a vessel containing the granular material. Then the particles are released one by one, beginning with the lowest one (with respect to the direction of gravity). Each particle moves until it reaches the next local minimum of the potential energy when it hits either the wall or other previously released grains. When a particle reached its local minimum it stays in that position, i.e. a falling particle cannot cause the motion of another previously released particle. This behavior means that inertia of the moving particles is neglected as soon as they hit another particle or the wall, and hence it corresponds to the case of very large damping. The advantage of this algorithm is its numerical simplicity, i.e. it is possible to investigate much larger systems than using molecular dynamics. Recently it has been applied in simulations of several problems as size segregation [41] and the motion of particles in a rotating cylinder [42,43]. Obviously the simplification of infinite damping is not valid in each case, and hence one has to carefully investigate whether it is justified to apply this algorithm (see e.g. [45]).

There are some more simulation methods for the dynamics of granular material which we want to mention only: Peng and Herrmann applied a Lattice Gas Automaton to investigate density waves in a pipe [46]. Caram and Hong investigated granular flows using a Random Walk technique [47].

There are some models to simulate complex shaped grains which are composed of spheres. In the model by Gallas and Sokołowski [48] the grains consist of two spheres connected with each other by a stiff bar. Walton and Braun applied more complicated particles consisting of four or eight spheres rigidly connected with each other [49]. Using this model they examined the transition from stationary to sliding flow and the transition from sliding to raining flow in a rotating drum. Pöschel and Buchholtz [50,38] describe grains built up of five spheres where one of them is located in the center of the grain and four identical spheres are at the corners of a square. Each pair of neighboring spheres is connected by a damped spring. The latter model was applied to the rotating cylinder [29] and it was shown that the simulation results agree much better with the experiment than equivalent calculations with spheres. Especially it was shown that one can reproduce stick–slip motion and avalanches which was not possible in simulations using spheres. The inclination of the surface of the material in the rotating cylinder and the dependency of the inclination on the angular velocity, however, did not agree well with the experiment.

Mustoe and DePorter [51] proposed a particle model where the boundary geometry of the particles is defined (in local coordinates) by

$$f_i(x_i, y_i) = \left[\frac{|x_i|}{a_i}\right]^{n_i} + \left[\frac{|y_i|}{b_i}\right]^{n_i} - 1 = 0. \qquad (7)$$

Varying $n_i$ between 2 and $\infty$ the shape of the $i$th particles varies continuously from elliptical to rectangular. To detect contacts of pairs of particles described by eq. (7) one has to solve numerically equations of high order for each pair of possibly touching particles in each iteration step. This requires an iteration technique which converges slowly for higher exponents $n_i$ and which makes hence the algorithm numerically complicated. Hogue and Newland [52] investigated the flow of granular material on an inclined chute and through a hopper using a model where the boundaries of the convex particles are given by polygons with up to 24 vertices each. To detect whether two particles touch each other one has to calculate the intersections between each pair of vertices. During collisions energy is dissipated according to Stronge's energy dissipation hypothesis [53] in normal direction, whilst Coulomb's friction law models the energy losses also in tangential direction.

Using grains which consist of interconnected spheres or of particles described by eq. (7) it is not possible to simulate particles with sharply formed corners. For some effects it seems to be essential to simulate such particles to reproduce the experimental observed effects. This point is discussed in detail in [38]. Tillemans and Herrmann [54,55] proposed a two dimensional particle model where the particles are convex polygons. When two particles $i$ and $j$ touch each other, i.e. when there is an overlap, there acts the force

$$\vec{F}_{ij} = F_{ij}^N \; \vec{e}^N + F_{ij}^T \; \vec{e}^T \tag{8a}$$

$$F_{ij}^N = \frac{Y\;A}{L_c} - m_{ij}^{eff} \; \gamma_N \; \left( \vec{v}^{rel} \cdot \vec{e}^N \right) \tag{8b}$$

$$F_{ij}^T = -\min \left( m_{ij}^{eff} \gamma_T \left| \vec{v}_{rel} \cdot \vec{e}^T \right|, \mu \left| F_{ij}^N \right| \right) \; . \tag{8c}$$

$A$ is the compression area of the particles (overlap area), Y is the Young module of the material. The effective mass $m_{ij}^{eff}$ is given by eq. (6e), the relative particle velocity is $\vec{v}^{rel} = \dot{\vec{r}}_i - \dot{\vec{r}}_j$, where $\vec{r}_i$ points to the center of mass of the particle $i$, and $L_c$ is the characteristic size of the particle. The unit vectors are in the direction of the line that connects the intersection points of the overlapping particles $i$ and $j$ ($\vec{e}^T$) and perpendicular to this line ($\vec{e}^N$). Only convex particles are allowed. This model was used e.g. in simulations of shear cells, earth quakes and flow through hoppers [55]. The results differ significantly from similar calculations using spheres. For the case of the hopper simulation they found clogging and arching which could not be found in simulations with spheres. A similar model was used earlier by Handley [56] who investigated the fracture behavior of brittle granular material. Potapov et al. [57,58] investigated a similar model for solid fracture. Initially they subdivide a macroscopic two dimensional body into many small equilateral triangles (elements). The forces in the body are resolved as forces at inter–element contacts of two different types which they call "glued" and "collisional". Glued contacts are the joints between elements interior to a solid body which can support stresses. When the tensile stresses exceed a certain value, cracks form and the glued bond breaks. Collisional contacts are contacts between the surface elements of the body (which eventually collides with other bodies or walls), and contacts between triangles in the bulk of the body where glued contacts have been broken. In the three latter models [54–58] the convex grains have been Voronoi polyhedra in two dimensions [59].

In this paper we present a new particle model for molecular dynamics of granular material in two dimensions where the particles are simulated using a similar model which has some advantages. In the following section we describe the model, the forces acting between the grains are derived in sections III and IV. Some aspects of the implementation and the performance of the algorithm are discussed in section V and some sample results based on this particle interaction model will be briefly discussed in section VI.

## II. THE MODEL

In our model the grains are composed of an arbitrary number of ideal elastic triangles which are connected by beams (see fig. 1). A beam in our sense is an deformable damped bar which is subjected to forces in the direction of its axis and transverse to its axis, i.e. to normal and shear forces, and to moments acting on its ends. Triangles which belong to the same grain do not interact with each other. When two triangles of different grains collide, i.e. if there is an overlap between both, they feel a restoring force. Hence the force acting upon the triangle $i$ belonging to the grain $j$ ($j \in \{1, 2, \ldots, N\}$) is

$$\vec{F}_i^j = \sum_{l=1, l \neq j}^{N} \sum_{k=1}^{n_t(l)} \vec{\Gamma}_{ik}^{jl} + \sum_{k=1}^{n_b^j(i)} \vec{\Lambda}_{ik}^j + \vec{\Phi}(\vec{r}_i^j, \dot{\vec{r}}_i^j) \;. \tag{9}$$

The first sum in the first term in eq. (9) runs over all grains $l$ except the $j$th, the second sum runs over all triangles $k \in \{1, 2, \ldots, n_t(l)\}$, the $l$th grain consists of. $\vec{\Gamma}_{ik}^{jl}$ is the force which acts between the triangles $i$ and $k$ which belong to different grains $j$ and $l$ (see section III). The sum in the second term runs over all beams connecting the triangle $i$ of the grain $j$ with other triangles of the same grain. $\vec{\Lambda}_{ik}^j$ is the force induced by the beam $k$ acting on the $i$th triangle of the grain $j$. They originate from the distortion of the beam that connects the triangle $i$ with another one of the same grain. The acting forces and momenta will be discussed below in section IV. A similar model for the beam forces has been introduced earlier by Herrmann et al. for the fracture of disordered lattices [60]. The last term $\vec{\Phi}(\vec{r}_i^j, \dot{\vec{r}}_i^j)$ describes the action of an external force, as e.g. gravity, on the triangles.

During its deformation a beam dissipates energy similar to a linearly damped spring, i.e. proportional to its deformation rate (see section IV). To avoid time consuming calculations due to Steiner's law each beam is fixed at both ends in the center of mass points of the triangles it connects. When the beam is not deformed by eventually applied forces (the beam is at rest) it is perpendicular to the neighboring edges of the triangles which it connects. Fig. 1 shows some examples of grains. The only restriction concerning the number, the shape and the position of the triangles is caused by the condition that the beams are fixed in the center of mass.

The next section III gives a detailed description of the force that acts when triangles collide, in section IV the forces due to deformed beams are derived.

### III. THE INTERACTION OF THE TRIANGLES

In this section we describe the calculation of the term $\vec{\Gamma}_{ik}^{jl}$ in eq. (9) originating from the compression of two triangles $i$ and $k$ which belong to the grains $j$ and $l$, respectively. When the triangles are compressed there exists a virtual "overlap area" which leads to an elastic restoring force $\vec{\Gamma}_{ik}$. For simplicity here and in the following we drop the upper indexes of the variables. Areas of triangles $XYZ$ and quadrangles $WXYZ$ we denote by $\Delta(XYZ)$ and $\square(WXYZ)$. Fig. 2 shows the most frequently found type of a collision between the triangles $A_iB_iC_i$ and $A_kB_kC_k$, i.e. a corner of one triangle deforms a side of another one. The absolute value of the restoring force $\vec{\Gamma}_{ik}$ is given by the shadowed area $\Delta(S_1S_2A_k)$ of the triangle $S_1S_2A_k$ times the Young module $Y$. Its direction is perpendicular to the intersection line $\overline{S_1S_2}$ (Poisson hypothesis, see e.g. [23]). Hence resulting momenta acting upon the triangles $A_iB_iC_i$ and $A_kB_kC_k$ read

$$\vec{M}(A_iB_iC_i) = H\vec{G}_i \times \vec{\Gamma}_{ik} \tag{10a}$$

$$\vec{M}(A_kB_kC_k) = H\vec{G}_k \times \vec{\Gamma}_{ki} = -H\vec{G}_k \times \vec{\Gamma}_{ik} \ . \tag{10b}$$

The vector $H\vec{G}_i$ points from the middle point of the line $\overline{S_1S_2}$ to the center of mass of the triangle $A_iB_iC_i$.

Although the case shown in fig. 2 is the mostly occurring type of collisions there are several other types which treatment has to be discussed below.

Classifying the possible interactions of the triangles we define five different types of collisions which are slightly different from those in [57]. The detection of overlapping polygons is a very important problem in computer graphics too, where one frequently has to decide whether two objects cover each other, and which of them is in the foreground or in the background, respectively. Therefore we found it helpful and inspiring first to have a look into advanced methods in computer graphics, e.g. [61,62]. In all figures 2–7 the overlapping area is drawn extremely exaggerated. Typically the overlapping area of two interacting triangles does not exceed a few tenth of a percent of the area of the triangles.

1. For the first type (fig. 3) there are two intersection points $S_1$ and $S_2$ lying at the same edge of one of the interacting triangles. One of the situations drawn in fig. 3 corresponds to the standard collision (fig. 2). We calculate the area $\Delta(S_1 S_2 A_i)$ (shadowed area) given by the intersection points $S_1$ and $S_2$ and the included point $A_i$. Because the overlapping area is definitely smaller than half the area of the triangles the force is proportional to the minimum between $\Delta(S_1 S_2 A_i)$ and $\Delta(A_i B_i C_i) - \Delta(S_1 S_2 A_i)$

$$\left|\vec{\Gamma}_{ik}\right| = \Gamma_{ik} = Y \cdot \min\left\{\Delta(S_1 S_2 A_i), \Delta(A_i B_i C_i) - \Delta(S_1 S_2 A_i)\right\} . \tag{11}$$

   The force acts perpendicular to the line between the intersection points $\overline{S_1 S_2}$.

2. The second type (fig. 4) is characterised by two intersection points lying on different edges for both triangles. We calculate the areas $\Delta(S_1 S_2 A_i)$ (dark gray) and $\Delta(S_1 S_2 A_k)$ (light gray). The force is proportional to the overlap

$$\left|\vec{\Gamma}_{ik}\right| = \Gamma_{ik} = Y \cdot [\min\{\Delta(S_1 S_2 A_i), \Delta(A_i B_i C_i) - \Delta(S_1 S_2 A_i)\} +$$
$$\min\{\Delta(S_1 S_2 A_k), \Delta(A_k B_k C_k) - \Delta(S_1 S_2 A_k)\}] . \tag{12}$$

   It acts perpendicular to the line between the intersection points as in the previous case.

3. The third type (fig. 5) is characterised by four intersection points, each pair of two points lies on one edge of both triangles. Therefore the third edge of both interacting triangles does not intersect the edges of the other triangle. There are two pairs of forces proportional to the quadrangular area $\square(S_1S_2S_3S_4)$ of the overlap

$$\left|\vec{\Gamma}_{ik}\right| = \Gamma_{ik} = \frac{1}{2} \cdot Y \cdot \square(S_1S_2S_3S_4) \ , \tag{13}$$

one acting perpendicular to the line $\overline{S_1S_2}$ and the other perpendicular to the line $\overline{S_1S_4}$. The pre–factor 0.5 is due to two pairs of acting forces instead of one for the first two collision types.

4. The fourth type (fig. 6) is characterised by four intersection points too, but here one of the triangles has intersection points at all its edges. Because interactions of this type occur extremely seldomly we did not implement an exact calculation for this case, but we calculate two interactions of type 1 with the intersection points $S_1$–$S_2$ or $S_3$–$S_4$ respectively, instead of solving the correct problem. In real simulation the fraction of this type occurs approximately $10^{-5}$ of the number of collisions. Nevertheless one has to deal with these extremely seldom events since otherwise they might cause problems from which the system cannot recover. If one does not care about these events usually the system gains in a few time steps after the event occurs a huge amount of kinetic energy, i.e. the system explodes.

5. The fifth and last type of interaction (fig. 7) is characterised by six intersection points $S_1 - S_6$. Equivalent to the forth type we do not calculate the exact interaction but substitute this calculation by solving three interactions of the first type with pairs of interaction points $S_1$–$S_2$, $S_3$–$S_4$ and $S_5$–$S_6$.

According to the forces $\vec{\Gamma}_{ik}$ we derive forces in parallel to the axes of the Cartesian coordinate system and moments $\vec{M}$ acting with respect to the center of mass points $G_i$ and $G_k$ as described in equations (10).

# IV. STRESSES IN BEAMS

When torques and forces are applied to a beam the beam deforms. Since all deformations are assumed to be small compared with the size of the beam one can superpose the deformations originating from different applying forces and torques [63,64]. Fig. 8 shows an (infinitesimally) deformed beam with radius of curvature $\rho$. We find the approximation

$$\tan\left(\frac{d\Theta}{2}\right) = \frac{1}{2 \cdot \rho} \cdot dx \quad \text{or} \quad \rho = \frac{dx}{d\Theta} \tag{14}$$

with $\tan(d\Theta) = d\Theta$.

The dashed line $\overline{ss}$ is the neutral surface length of which does not change during the deformation. All other fibres below and above the neutral surface are in tension or in compression, respectively. The length of a fibre in distance $y$ from the neutral surface is $(\rho + y)\,d\Theta = (1 + y/\rho)\,dx$ and hence the strain reads $\epsilon = y/\rho$. With Hooke's law $\sigma = E \cdot \epsilon$ one finds $\sigma = Ey/\rho$. Finally the resulting momentum $M$ is

$$M = \int_A \sigma y \, dA = \frac{EI}{\rho} \tag{15}$$

where the moment of inertia $I = \int y^2 dA$ depends on the geometrical shape of the cross section of the beam. In the present paper we assume that the beams have quadratic cross section of width $b$ and hence $I = b^4/12$. When we express the radius of curvature $\rho$ by

$$\frac{1}{\rho} = \frac{d^2 v}{dx^2}, \tag{16}$$

where $v$ is the deflection of the beam from its initial position (with the approximation $\tan\Theta \approx \Theta$). From eqs. (15) and (16) one finds the basic equation for the deflection of a beam:

$$\frac{d^2 v}{dx^2} = v'' = +\frac{M}{E\,I} \tag{17}$$

This equation has to be solved to find the forces and moments which act when a beam is deformed. The integration constants are used to satisfy the boundary conditions.

There are three different deformation modes for beams: elongation, shearing and bending. In the following three subsections IV A, IV B and IV C we will discuss the deformation rules in detail.

### A. Elongation of beams

Since the exact notation of the acting forces and moments as vector functions of the coordinates of the triangles the beam connects, is not very instructive but confusing due to the length of the formulae expressions we discuss for simplicity of notation the deformation of the beams here and in the following in local coordinates. Instead of providing the exact vector notation we rather discuss absolute values. In all cases the directions in which the forces and moments act are very clear, moreover they are given for each case explicitly in the figures 9–11 and in figure 14. For the implementation of the algorithm, however, one should note that it is necessary to transform the given expressions for the forces and momenta into the coordinate system of the triangles. These transformations require a considerable part of the computation time.

Fig. 9 displays the shape and the location of a beam of length $L$ at rest, i.e. when no deforming forces $F_A$, $F_B$ or moments $M_A$, $M_B$ at the ends $A$ and $B$ of the beam apply. The $x$–axis is drawn with a thin solid line, i.e. at rest the beam lies on the $x$–axis.

For the elongation deformation we find according to the linear Hooke law the restoring force

$$\left|F^{el}\right| = |\Delta L| \cdot E. \tag{18}$$

### B. Shearing

We want to calculate the moments and forces $M_A$, $M_B$, $F_A$ and $F_B$ of a sheared beam as drawn in Fig. 10. For a beam of length $L/2$ which is fixed at one end (at $x = 0$) and where acts the shear force $F$ (in negative $y$–direction) at the other end (at $x = L/2$) we find for the local moment $M(x) = F\left(\frac{L}{2} - x\right)$ and hence eq. (17) reads

$$v'' = \frac{F}{EI}\left(\frac{L}{2} - x\right) \tag{19}$$

with the boundary conditions $v(0) = 0$ and $v'(0) = 0$. We find for the vertical deviation $\Delta/2$ at the point $x = L/2$

$$\Delta/2 = \frac{FL^3}{24\,EI} \tag{20}$$

force which acts at the free end of the beam.

One can assume that the sheared beam in Fig. 10 is built up of two of such beams of length $L/2$ as described above. Finally one finds

$$\Delta = \frac{FL^3}{12\,EI} \tag{21}$$

and hence

$$F_A = -F_B = F = \frac{12EI}{L^3}\Delta \tag{22a}$$

$$M_A = M_B = F \cdot \frac{L}{2} = \frac{6EI}{L^2}\Delta\,. \tag{22b}$$

### C. Bending

When a beam undergoes bending deformation (fig. 11), the resulting forces and moments can be found by superposing the forces and moments that would act if the beam would be bent at each side separately. Therefore we can restrict ourself on calculating the forces and moments according to the single sided deformation drawn in fig. 12.

The deformation in fig. 12, however, can be understood as a deformation of a beam with free ends according to a single sided moment $M$ (fig. 13) superposed by a moment $M_b$ (fig. 12) that assures that the angle $\Theta_B$ (fig. 12) results to zero.

First we consider one sided bending of the beam (fig. 12). For a beam as drawn in fig. 13 where the moment $M$ acts at one of its both sides one finds easily

$$\Theta_A^* = \frac{ML}{3EI} \tag{23a}$$

$$\Theta_B^* = -\frac{ML}{6EI}. \tag{23b}$$

Since the beam in Fig. 12 has the angle $\Theta_B = 0$ we find the moments acting at the ends of the beam in fig. 12 by superposing the moment $M_B$, which causes a virtual angle $\Theta_B^{**}$, fulfilling the condition $\Theta_B^{**} + \Theta_B^* = 0$:

$$M_B = \frac{3EI}{L}(-\Theta_B^*) = \frac{M}{2}. \tag{24}$$

The moment $M_B$ causes the angle

$$\Theta_A^{**} = -\frac{\frac{M}{2}L}{6EI} = -\frac{ML}{12EI} \tag{25}$$

and the resulting angle $\Theta_A$ (fig. 12) is

$$\Theta_A = \Theta_A^* + \Theta_A^{**} = \frac{M_A L}{4EI}. \tag{26}$$

Hence, finally we find for the different moments and forces for the bending deformation (fig. 12)

$$M_A = -\frac{4EI}{L}\Theta_A \tag{27a}$$

$$M_B = -\frac{2EI}{L}\Theta_A \tag{27b}$$

and with

$$M_A = -F_B L - M_B \tag{28}$$

$$F_B = -F_A = \frac{6EI}{L^2}\Theta_A. \tag{29}$$

(Note that the moments $M_A$ and $M_B$ in eq. (28) have to be added as if they would apply to the same end of the beam.)

If the beam is bent at both ends by angles $\Theta_A$ and $\Theta_B$ (Fig. 11) the resulting moments $M_A$ and $M_B$ and forces $F_A$ and $F_B$ can be calculated by a superposition of the forces and moments according to two independent bending deformations of the type discussed in the previous case.

The angle $\Theta_A$ generates the moments $M_A^*$ and $M_B^*$ according to eqs. (27)

$$M_A^* = -\frac{4EI}{L}\Theta_A \tag{30a}$$

$$M_B^* = -\frac{2EI}{L}\Theta_A \tag{30b}$$

and the angle $\Theta_B$ causes the moments

$$M_A^{**} = -\frac{2EI}{L}\Theta_B \tag{31a}$$

$$M_B^{**} = -\frac{4EI}{L}\Theta_B . \tag{31b}$$

Hence we find the resulting moments

$$M_A = M_A^* + M_A^{**} = -\left(\frac{4EI}{L}\Theta_A + \frac{2EI}{L}\Theta_B\right) \tag{32a}$$

$$M_B = M_B^* + M_B^{**} = -\left(\frac{2EI}{L}\Theta_A + \frac{4EI}{L}\Theta_B\right), \tag{32b}$$

and with $M_A + M_B = -F_B L$ the resulting forces

$$F_B = -F_A = \frac{6EI}{L^2}\left(\Theta_A + \Theta_B\right) . \tag{33}$$

Comparing the results from sections IV B and IV C one can show that each deformation of a beam can be expressed by bending and elongation, i.e. the shear deformation need not to be considered. The proof is given in Appendix A.

The deformation of the beams is damped proportionally to the deformation rates

$$M_A^{(d)} = -\frac{\gamma\, I}{L}\,\dot{\Theta}_A, \tag{34a}$$

$$M_B^{(d)} = -\frac{\gamma\, I}{L}\,\dot{\Theta}_B, \tag{34b}$$

$$F_A^{N\,(d)} = -\gamma\,(\vec{v}_A - \vec{v}_B)\cdot\overrightarrow{AB}, \tag{34c}$$

$$F_B^{N\,(d)} = \gamma\,(\vec{v}_A - \vec{v}_B)\cdot\overrightarrow{AB}, \tag{34d}$$

where $\gamma$ is the damping coefficient of the beam material. As proven in the appendix the deformation of the beam is in linear approximation completely determined by the angles $\Theta_A$ and $\Theta_B$ and the length $L$. Hence there is no damping force acting in shear direction.

## V. IMPLEMENTATION OF THE ALGORITHM

We implemented a molecular dynamics algorithm using the particle model described in the previous sections in FORTRAN. A Gear predictor corrector scheme of fourth order [65,66] was applied to integrate Newton's equations of motion. We have to proceed both the forces caused by the elastic deformation of the triangles as well as the forces induced by the beams for each iteration step (eq. 9). Since every particle can interact with each other one causing a high algorithmic complexity, we applied a Verlet neighborhood list [67] to decrease the amount of computer time. As discussed below (see table I) the calculation of the triangle intersections is the most time intensive part of the calculation. Hence we applied an advanced version of the neighborhood list method to keep the fraction of the computation of the triangle–interactions as small as possible.

Our list method acts in two steps. In the first step we prepare neighborhood lists for each triangle to reduce the number of possible interactions which has to checked. Because we have to investigate all possible triangle interactions the time for this step rises as the square of the total number of triangles $T \sim (N \cdot n_t)^2$. (For simplicity we assumed for this estimate that all $N$ grains consist of $n_t$ triangles each. Hence there are $(N \cdot n_t) \cdot ((N-1) \cdot n_t)$ possible triangle–interactions.) In the second step we check for each entry of the constructed neighborhood list whether the corresponding triangles interact indeed or not. If they interact, i.e. if there is an overlap area, the type of interaction is determined according to the classification scheme in section III.

Finally we construct three lists for the interactions of the first three types of the classification scheme. The interactions of the forth and fifth type are replaced by interactions of the first type as described above. The computer time for this step rises linearly with the number of particles $N$ and with the number of triangles $n_t$ belonging to each particle. Different from usual Verlet tables these lists contain only triangles which definitely do interact, and hence we call the most time consuming subroutines for the calculation of the forces only for triangles which do interact in one of the five manners described above (section III).

The calculation of the forces induced by the beam deformations follows straight forward the formulas described in section IV.

In table I we present a detailed performance analysis of the algorithm. The numbers in the table refer to simulations on a DEC 3000/700 workstation. As visible from the data most of the time (approximately 70%) is necessary to construct the interaction list, because we check here every pair of neighbored triangles for intersections. The construction of this lists give the possibility to classify the interactions and therefore to use for each kind of interaction its own optimised algorithm. This prevents us from calculating useless intersection points, areas and other data.

The computer time needed for the predictor–corrector integration is very low (3.7%). Thus it seems to be not useful to optimise the integration procedure, or to decrease the accuracy of the integration to safe computer time. The calculation of the beam forces is not time expensive as well.

From the data in table I it is obvious that triangle interactions of the first type dominate. Approximately 98% of the occurring interactions are of this type. Interactions of the fourth and fifth type, where our program yields approximative results only, occur extremely seldomly. We found them only a few times during all our simulations.

## VI. FIRST RESULTS

### A. The collision of a two particles

To demonstrate the non–trivial behavior of colliding non–spherical particles and to check the correctness of our implemented model we want to present the results of an experiment where a quadratic grain collides with another resting one in the absence of gravity. The grains consist of four equal triangles each, forming a simple quadratic grain as shown in fig. 1. The parameters of the materials are: $Y = 2 \cdot 10^7 \, g/(cm \, sec^2)$, $E = 1 \cdot 10^5 \, g/(sec^2)$, $I = 1 \cdot 10^{-4} cm^3$ and $\gamma = 9 \, g/sec$. The value of the Young–module $Y$ agrees with most of the simulations of spheres which can be found in the literature.

Figs. 15 (left figures) show the time evolution of the particles as a sequence of stroboscopic snapshots for different relative velocities of the particles a) $v_{rel} = 10 \, cm/sec$, b) $v_{rel} = 50 \, cm/sec$ and c) $v_{rel} = 100 \, cm/sec$. One of the triangles of each grain has been filled to visualise the rotation of the grains. Initially the velocity of the left particle is $\vec{v} = (v_{rel}, 0) \, cm/sec$ and the right particle rests. For low velocities $v < v_{tc} \approx 70 cm/sec$ the grains collide twice within a very short time. Therefore the traces of the particles that undergo soft collisions with velocity values $v < v_{tc}$ differ qualitatively from the traces of particles colliding with higher relative velocity. Since it is hard to display the complicated motion of the grains in detail we refer to our World Wide Web–site URL http://summa.physik.hu-berlin.de:80/~thorsten/MDASGP.html [68] where one can find animated sequences of various problems mentioned in the current paper. The right hand side in fig. 15 shows the rotational and kinetic energies and the total energy as a function of time. The initial kinetic energy of the translational motion of the left particle results in kinetic energy of both particles due to translational motion and rotation. A part of the mechanical energy is lost due to dissipation in the beams. The relative amount of rotational energy depends strongly on the impact velocity as expected from the discussion above. The dependence of the extinction of the rotational degree of freedom on the impact rate is demonstrated in fig. 16 too. This figure shows the time evolution of the angular velocity of the grains $\omega = \frac{1}{n_t} \left| \sum_{i=1}^{n_t} (\vec{v}_i - \vec{v}_g) \times \vec{s}_i \right|$ over time for different values of the impact velocity. $\vec{v}_g = \frac{1}{n_t} \sum_{i=1}^{n_t} \vec{v}_i$ is the velocity of the grain and $\vec{s}_i$ is the vector pointing from the center of mass point of the grain to the center of mass point of the triangle $i$. $n_t$ is the number of triangles the grain consists of. For high velocity one observes an abrupt change of the rotational motion at the time of the collision and a short damped oscillation according to the excited inner degrees of freedom of the grain (distortion of beams) and their relaxation. After the collision the particles move on with constant translational and rotational velocities. For low impact rate we find a quite different behavior: First the particles collide as in the previous case, resulting in a certain rotational motion. Short time after the first collision, however, they collide a second time with different edges. Finally we find for the latter case of slow collisions a very small resulting angular velocity because the second collision causes angular moments in opposite directions, hence the resulting moment is small.

Observing the motion of a moving particle colliding with a resting one (see [68]), i.e. the simplest possible contact of particles, one can remark that even for this system there is a complex behavior. For rigid spheres the velocities after a collision as a function of the impact velocities (eq. (1)) can be calculated analytically solving the viscoelastic equations of the spheres [8]. We guess that this will not be possible for the case of our particles.

### B. Outflow of a hopper

The outflow of a hopper is of high interest not only because of the technological importance of this process, but also because of the exciting phenomena as clogging and density waves which have been discovered recently (e.g. [69]). Density waves and clogging have been investigated using molecular dynamics in two and three dimensions in various papers, e.g. [23,70,71]. There is at least one other interesting and surprising phenomenon, detected recently by Evesque and Meftah [72,73]: They found that a hour glass "ticks" much slower if it is subjected to vertical vibrations $y = A\cos(\omega t)$. The effect seems to depend on the frequency $f = \omega/2\pi$ and on the acceleration of the vibration $\Gamma = A\omega^2$. For frequencies between $40\,Hz \leq f \leq 60\,Hz$ surprisingly they observed for some accelerations that the flow almost stops.

We investigated this effect using different particle models and we found that the effect could be reproduced with the new particle model only. The results will be described in detail in a forthcoming paper [74].

Fig. (17) shows snapshots of the system with $N = 400$ complex particles. For an animated sequence we refer to our WWW–site [68]. The left figure shows the outflowing hopper, the flow varies irregularly and the grey scale codes for the particle velocities. When animating the figures one clearly observes density waves [68]. The right hand figure shows the same system a few moments later. The flow has stopped due to clogging.

## C. Granular flow in a rotating cylinder

The flow of granular material in a rotating cylinder is one of the most popular problems in the field of granular materials. It has been investigated experimentally and theoretically by many authors using various techniques (see e.g. [75–78,27,43,50] and many others). The results shall not be discussed here. We applied the algorithm to this problem too. For the detailed description of the results see [79]. Here we only want to demonstrate the abilities of the method and to mention the main results briefly.

Fig. 18 shows snapshots of the simulation of a slowly rotating cylinder. The grey scale codes for the velocity of the grains, black means high velocity, black codes for high velocity. For low angular velocity of the driven cylinder one finds experimentally stick–slip flow (e.g. [75]), i.e. the material moves downwards not homogeneously but it forms avalanches. In the right figure one observes an avalanche on the top of the material indexed by light grey shadowed grains. The left figure shows the same system immediately before the avalanche. Avalanches are relatively seldom events. Most of the time the particles rest with respect to each other, i.e. they move only due to the rotation of the cylinder (left side of fig. 18). When increasing the angular velocity of the cylinder there is a relatively sharp transition between the stick–slip flow and the homogeneous regime. This transition could be found in the simulation using non–spherical grains [79]. Animated sequences for both regimes, stick–slip and continuous flow, can be accessed via World Wide Web [68].

Fig. 19 shows the dependence of the difference between the inclination of the material and the angle of repose on the angular velocity of the cylinder. The line displays the behavior found by Rajchenbach $\Theta - \Theta_c \sim \Omega^2$ [75]. The simulation data are in good agreement with this observation.

Both results mentioned, the sharp transition between the flow regimes and the inclination as a function of the angular velocity of the cylinder, agree with experimental observations. We have not been able to find these effects using simple spheres or using particles composed of spheres [50,29] in molecular dynamics simulations.

## VII. CONCLUSION

We presented a model for the simulation of the particles of a granular material. Each particle consists of triangles which are connected by deformable beams. There are no restrictions concerning the shape of the grains (convex or concave), nor the number or the shape of the triangles a grain consists of. To preserve calculation time it is favourable to chose arrangements of the triangles so that the beams are fixed at the center of mass points of the triangles. Otherwise one needs additional computation time caused by the Steiner law.

When two triangles of different grains collide there acts an elastic restoring force proportionally to the compression of the particles. During the collision of the triangles the energy is preserved. A collision of a triangle with another one causes moments and forces acting on this triangle and hence a deformation of the beams which connect the triangles to neighboring ones. The deformation of the beams causes forces and moments acting on the neighboring triangles.

Beams can be deformed in axial direction (elongation) and in shear direction and they can be bent. When a beam is deformed it dissipates energy proportionally to the deformation rate.

In the present paper we described beams that recover completely while dissipating mechanical energy when the deforming moments and forces vanish. Our model, however, is not restricted to this case. Interesting phenomena are plastic deformation and wear which can be easily simulated by introducing thresholds for the beam forces and moments or for the deformations. When a force or a moment or a deformation, respectively, exceeds the threshold the beam breaks or deforms permanently. Similar simulations with other models have been reported in the literature [55–57].

The proposed algorithm has been implemented in FORTRAN. We have demonstrated the behavior of a system of grains applying it to three examples of granular assemblies. In a simple collision simulation of two grains we discussed the trajectories of the grains as well as the evolution of the kinetic and rotational energies depending on the impact rate. Sample results for the flow out of a hopper and the motion of granular material in a rotating cylinder have been presented. In the latter case we found quantitative agreement of the simulation with the experiment that could not be found so far, neither with spherical grains nor with non–spherical grains of other type.

## ACKNOWLEDGMENTS

We thank Hans J. Herrmann, Stefan Schwarzer and Hans–Jürgen Tillemans for many helpful discussions and comments. This work has been supported by the Deutsche Forschungsgemeinschaft (grant Ro 548/5-1).

## APPENDIX A: SHEAR DEFORMATION CAN BE EXPRESSED BY BENDING

The deformation of a beam in the two dimensional space has three degrees of freedom: the distance of the relative positions of the ends of the beams $L = \sqrt{(x_A - x_B)^2 + (y_A - y_B)^2}$, and the angles $\Theta_A$ and $\Theta_B$. These values are the parameters in eqs. (32) and (33). The shear parameter $\Delta$ in eqs. (22) is a fourth parameter, i.e. one of them must be redundant. In the following we will show that shear deformation can be expressed by bending, at least in our approach of a linear deformation–force relation.

Suppose we have a deformed beam as drawn in Fig. 14 (bottom). Then applying eqs. (22) we find for the forces and moments according to shearing

$$M_A^s = M_B^s = \frac{6EI}{(L^*)^2} \Delta \tag{A1a}$$

$$F_A^s = -F_B^s = \frac{12EI}{(L^*)^3} \Delta \tag{A1b}$$

and according to bending (eqs. (32) and (33))

$$M_A^b = -\frac{2EI}{L^*} \Theta_B^* \tag{A2a}$$

$$M_B^b = -\frac{4EI}{L^*} \Theta_B^* \tag{A2b}$$

$$F_A^b = -F_B^b = -\frac{6EI}{(L^*)^2} \Theta_B^* \ . \tag{A2c}$$

The total forces and moments read

$$M_A^* = M_A^s + M_A^b = \frac{6EI}{(L^*)^2} \Delta - \frac{2EI}{L^*} \Theta_B^* \tag{A3a}$$

$$M_B^* = M_B^s + M_B^b = \frac{6EI}{(L^*)^2} \Delta - \frac{4EI}{L^*} \Theta_B^* \tag{A3b}$$

$$F_A^* = -F_B^* = \frac{12EI}{(L^*)^3} \Delta - \frac{6EI}{(L^*)^2} \Theta_B^* \ . \tag{A3c}$$

Turning the entire system by the angle $\Theta_A \approx \tan\Theta_A = -\frac{\Delta}{L^*}$ we obtain the new system as drawn in the top of fig. 14 with the new angles $\Theta_A = -\frac{\Delta}{L}$ and $\Theta_B = \Theta_B^* - \frac{\Delta}{L}$. With eqs. (32) and (33) we find

$$M_A = -\frac{4EI}{L}\Theta_A - \frac{2EI}{L}\Theta_B = \frac{6EI}{L^2}\Delta - \frac{2EI}{L}\Theta_B^* \tag{A4a}$$

$$M_B = -\frac{2EI}{L}\Theta_A - \frac{4EI}{L}\Theta_B = \frac{6EI}{L^2}\Delta - \frac{4EI}{L}\Theta_B^* \tag{A4b}$$

$$F_A = -F_B = -\frac{6EI}{L^2}(\Theta_A + \Theta_B) = \frac{12EI}{L^3}\Delta - \frac{6EI}{L^2}\Theta_B^* \tag{A4c}$$

Approximating $L$ by $\sqrt{L^2 + \Delta^2} = L^*$ evidently the results in eqs. (A4) coincide with eqs. (A3). Thus we do not need to consider shear deformation in our simulations since it is redundant here.

*Particle Technology*, Nisshin Engineering Co. Ltd. (Osaka, 1993).

TABLES

| Basic algorithm | portion of computation | subroutine | portion of computation |
|---|---|---|---|
| Construction of the lists | 73.7 % | Verlet list | 2.07 % |
|  |  | Classifying interaction | 70.62 % |
| Calculation of the triangle interaction | 7.87 % | TYPE 1 | 7.54 % |
|  |  | TYPE 2 | 0.24 % |
|  |  | TYPE 3 | 0.09 % |
| Calculation of the beam forces |  |  | 13.33 % |
| integration | 3.71 % | predictor | 1.16 % |
|  |  | corrector | 2.55 % |
| all others |  |  | 1.4 % |

TABLE I. The performance analysis of the algorithm. The calculation has been done on a DEC–3000/700 workstation (explanation in the text).

FIGURES

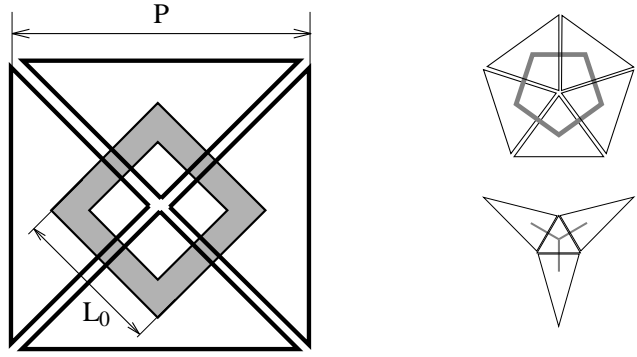

FIG. 1. Examples of grains composed of different numbers of triangles. The model is not restricted to convex grains.

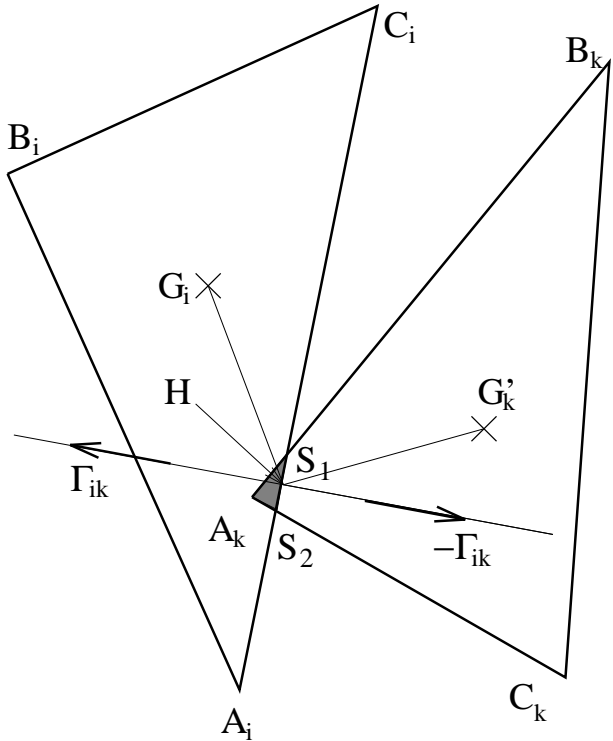

FIG. 2. The "standard" type of a collision between two triangles $A_iB_iC_i$ and $A_kB_kC_k$. The forces $\pm\vec{\Gamma}_{ik}$ are directed perpendicular to the intersection line $\overline{S_1S_2}$. Moments act with respect to the center of mass points $G_i$ and $G_j$. The absolute value of the interaction force $\left|\vec{\Gamma}_{ik}\right|$ is proportional to the shadowed intersection area $\Delta(S_1S_2A_k)$.

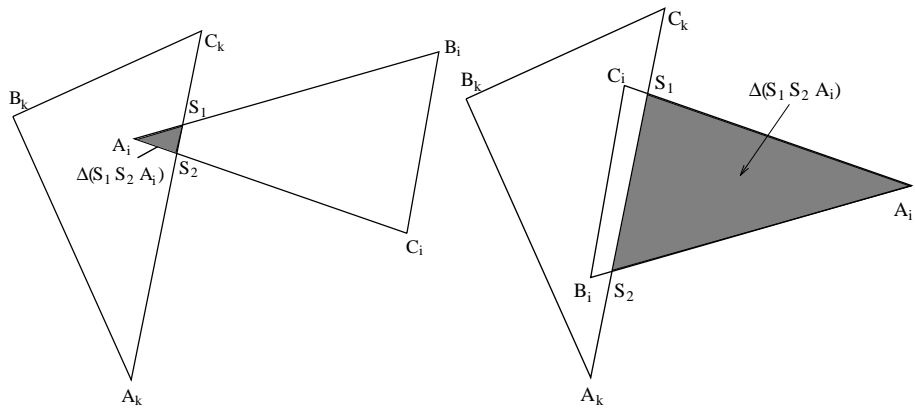

FIG. 3. The first type of collisions: There are two intersection points $S_1$ and $S_2$. Both lie at the same edge of one of the triangles.

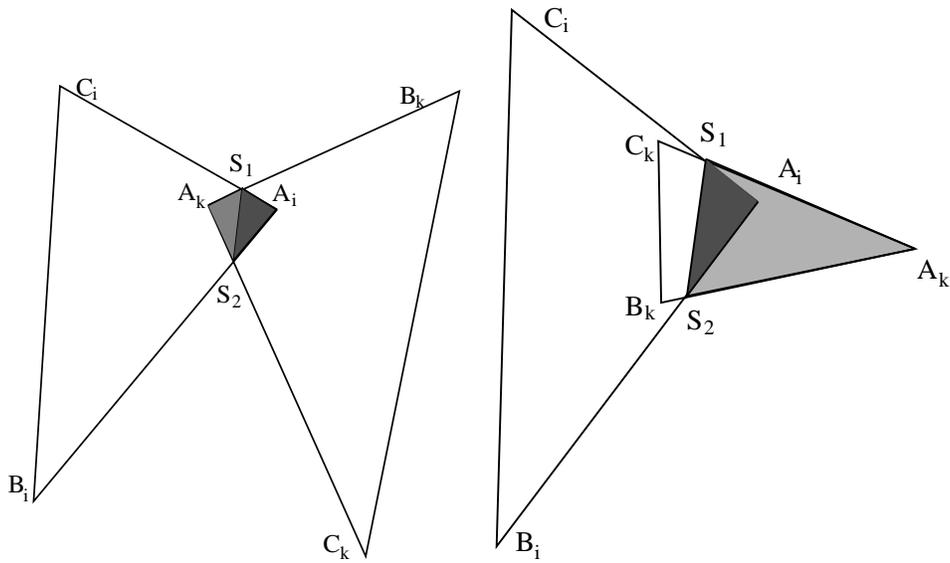

FIG. 4. The second type of collisions: There are two intersection points $S_1$ and $S_2$ lying on different edges for both triangles.

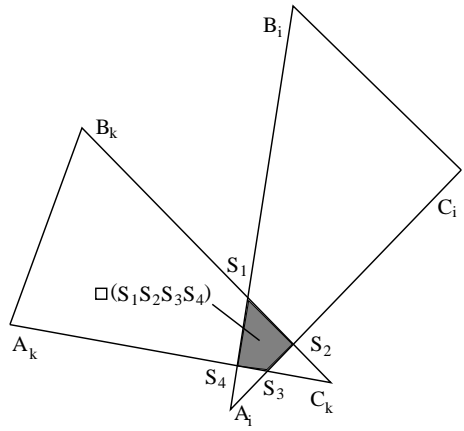

FIG. 5. The third type of collisions: There are four intersection points $S_1$, $S_2$, $S_3$ and $S_4$. Each pair of points lies on one edge of either the first or the second triangle.

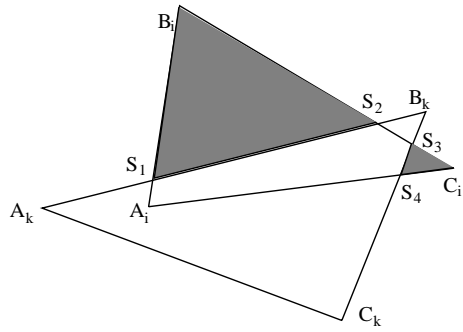

FIG. 6. The fourth type of collisions: There are four intersection points $S_1$, $S_2$, $S_3$ and $S_4$. One of the triangles intersects with all its three edges.

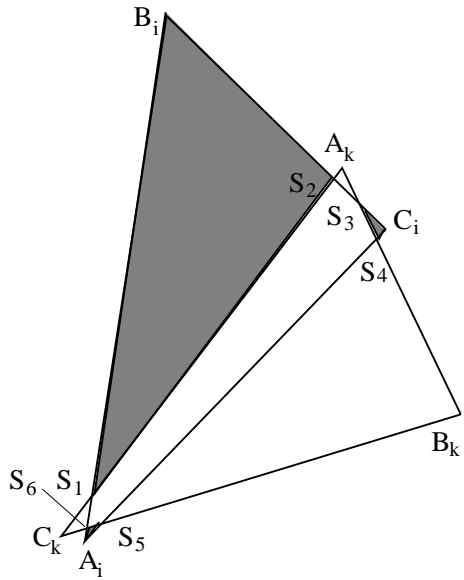

FIG. 7. The fifth type of collisions: There are six intersection points $S_1,\ldots,S_6$.

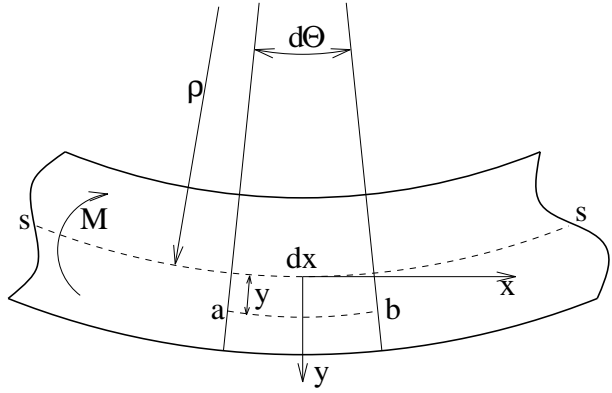

FIG. 8. An infinitesimally deformed beam with radius of curvature $\rho$. The fibres below or above the neutral fibre $\overline{SS}$ are in tension or in compression. For a beam with quadratic cross section $b^2$ the deformation leads to the resulting moment $M = \frac{E\,b^4}{12\,\rho}$.

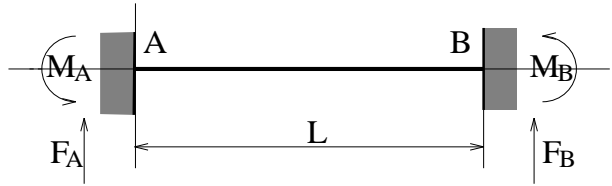

FIG. 9. The shape and the location of a beam at rest. Its length is assumed to be $L$, its ends $A$ and $B$ lie on the $x$–axis. When the beam is not deformed there do not act any forces $F_A$, $F_B$ or moments $M_A$, $M_B$. In the following figures 8-11 and in fig. 14 the $x$–axis is drawn as a thin horizontal line.

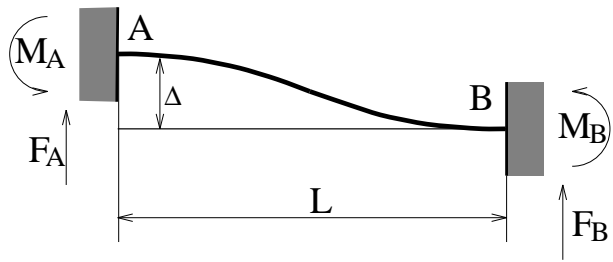

FIG. 10. When the beam in fig. 9 undergoes a shear deformation $\Delta$ there act the forces $F_A$ and $F_B$ and the moments $M_A$ and $M_B$. The forces and moments are given in eqs. (22).

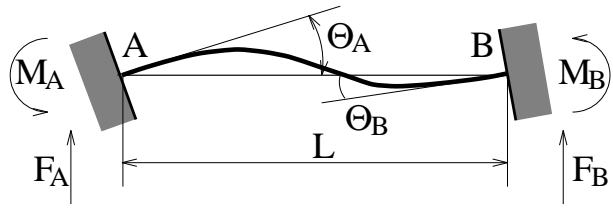

FIG. 11. A bent beam. The forces $F_A$ and $F_B$ and the moments $M_A$ and $M_B$ can be calculated by superposing the forces and moments of two single sided bent beams (fig. 12), one of them bent at the side $A$, the other one at the side $B$. The results are given in eqs. (32) and (33).

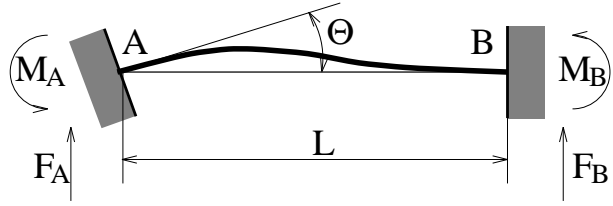

FIG. 12. A single sided bent beam. The forces and moments result from the superposition of the deformation drawn in fig. 13 and the restoring moment $M_B$ (fig. 12) which assures the angle $\Theta_B$ to be zero. The equations for the forces and moments are given in eqs. (27) and (29).

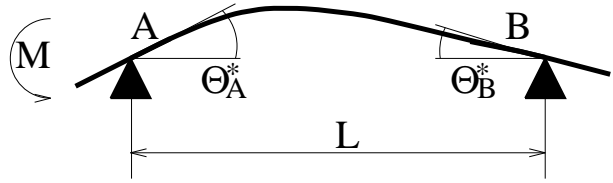

FIG. 13. If the ends of the beam are not fixed but freely moving, the applied moment $M$ causes bending due to the angles $\Theta_A$ and $\Theta_B$ given in eqs. (23).

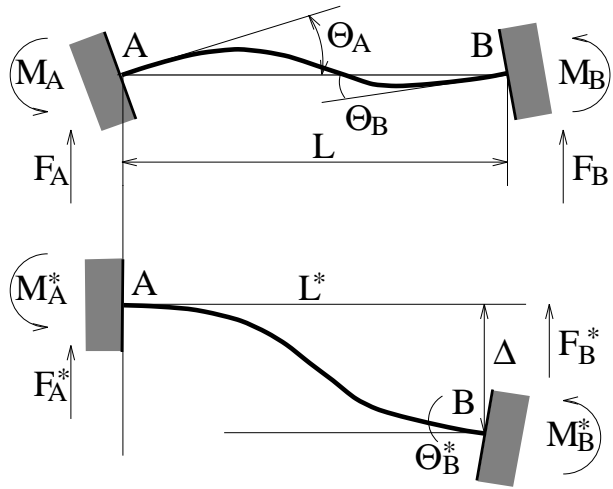

FIG. 14. The deformation of a beam in shear direction $\Delta$ (upper part) can be represented by bending deformation (lower part). The appendix contains the proof that one need not to consider shear *and* bending deformation in the linear approximation which is justified for small deformations.

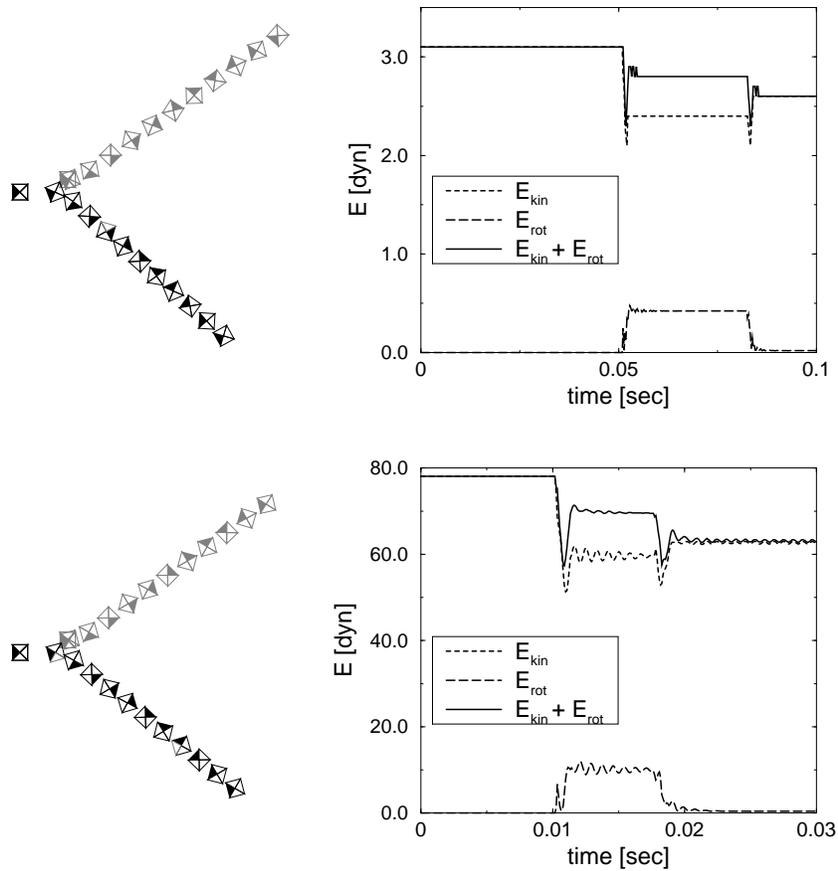

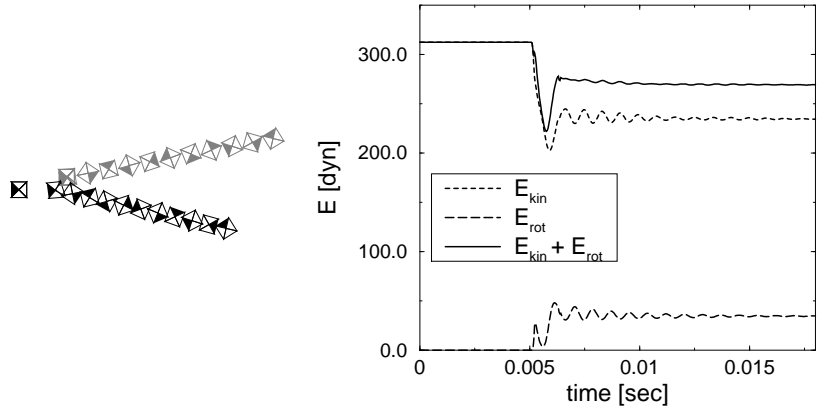

FIG. 15. The collision of a horizontally from left to right moving grain (black drawn) with a resting one (grey) for different impact rates $v_{rel} = 10\,cm/sec$ (top), $v_{rel} = 50\,cm/sec$ (middle), and $v_{rel} = 100\,cm/sec$ (bottom). The stroboscopic snapshots of the grains are take at equidistant time intervals. An animated sequence of the collisions can be accessed via World Wide Web [67]. The right hand figures display the kinetic, rotational and total energies of the particle collisions over time.

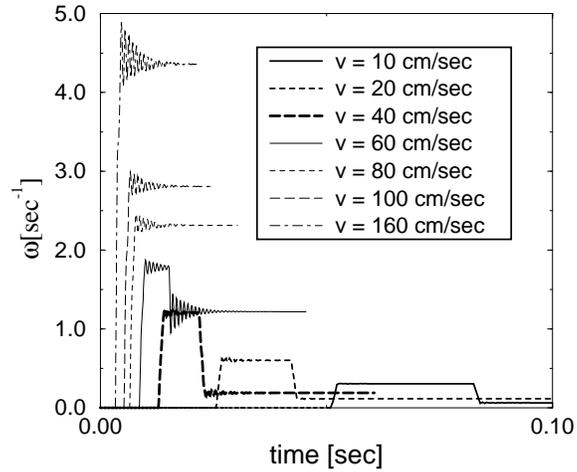

FIG. 16. The angular velocities of the particles over the time. For lower velocities the particles collide two times, therefore those collisions results in a small angular velocity of the grains.

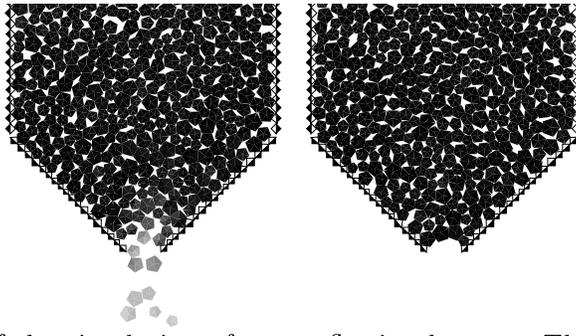

FIG. 17. *Snapshots of the simulation of an outflowing hopper. The left figure shows regular outflow, the right figure show a clogging situation. An animated sequence is available via World Wide Web [67].*

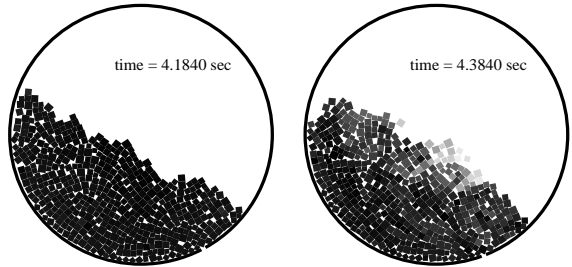

FIG. 18. *Snapshots of the simulation with $N = 500$ complex particles with $P \in [0.1\,cm; 0.2\,cm]$ in a rotating cylinder of diameter $D = 4\,cm$. The angular velocity is $\Omega = 0.1\,sec^{-1}$. The wall particles have not been drawn. The right snapshot shows an avalanche, the left one has been taken short time before. The grey scale codes for the particle velocity.*

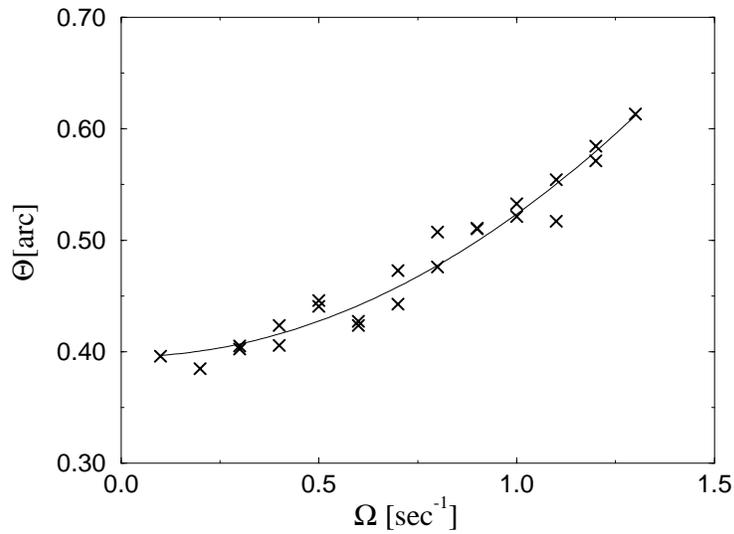

FIG. 19. *The inclination $\Theta$ of the material surface over the angular velocity $\Omega$. The dotted line displays the function which has been measured experimentally by Rajchenbach [73].*